\documentclass[conference]{IEEEtran}

\usepackage{graphicx}
\usepackage{color}
\usepackage{listings}
\usepackage{amssymb}
\usepackage{amsmath}
\usepackage{amsthm}

\usepackage{mathtools}
\usepackage{extpfeil}
\usepackage{enumerate}
\usepackage{proof}
\usepackage{url, hyperref}
\usepackage[normalem]{ulem}
\usepackage{xcolor, colortbl}
\usepackage{algorithm}
\usepackage{algorithmic}
\usepackage{comment}
\usepackage{xspace}
\usepackage{chngpage}
\usepackage{subfig}
\usepackage{centernot}
\usepackage{booktabs}
\usepackage{tabu}
\usepackage{tabularx}

\usepackage{wrapfig}

\usepackage{adjustbox}

\usepackage{pifont}

\newcommand\GG{\ensuremath{\square}}
\newcommand\FF{\ensuremath{\lozenge}}

\newcommand\code[1]{\lstinline|#1|}

\newcommand\eg{{\it e.g.,}\xspace}

\definecolor{light-gray}{gray}{0.80}

\newcommand{\rewrite}[3]{\ensuremath{#1} &\ensuremath{\vdash_E}& \ensuremath{#2} &\ensuremath{\leadsto #3}}
\newcommand{\inlrewrite}[3]{\ensuremath{#1 \vdash_E #2 \leadsto #3}}
\newcommand{\weaken}[3]{\ensuremath{#1} &\ensuremath{\vdash_S}& \ensuremath{#2} &\ensuremath{\leadsto #3}}
\newcommand{\inlweaken}[3]{\ensuremath{#1} \ensuremath{\vdash_S} \ensuremath{#2} \ensuremath{\leadsto #3}}

\newcommand\ignore[1]{}

\newcommand\added[1]{#1}

\newcommand*\OK{\ding{52}\xspace}
\newcommand*\TOUT{\textbf{T}\xspace}
\newcommand*\OOM{\textbf{M}\xspace}
\newcommand*\NOK{\ding{55}\xspace}

\newcommand*\UNK{{\textbf{?}\xspace}}

\newcommand{\header}[1]{\parbox{5em}{\centering #1}}

\begin{document}
\title{Source-Level Bitwise Branching for Temporal Verification}
%

\author{\IEEEauthorblockN{Yuandong Cyrus Liu,
Ton-Chanh Le, 
Eric Koskinen
}

  \IEEEauthorblockA{Stevens Institute of Technology, Hoboken, NJ 07030 USA}}

\maketitle              
\begin{abstract}
There is increasing interest in applying verification tools to programs that have bitvector operations. SMT solvers, which serve as a foundation for these tools, have thus increased support for bitvector reasoning through bit-blasting and linear arithmetic approximations. Still, verification tools are limited on termination and LTL verification of bitvector programs. 

In this \added{work}, we show that similar linear arithmetic approximation of bitvector operations can be done at the source level through transformations. Specifically, we introduce new paths that over-approximate bitvector operations with linear conditions/constraints, increasing branching but allowing us to better exploit the well-developed integer reasoning and interpolation of verification tools. We present two sets of rules, namely rewriting rules and weakening rules, that can be implemented as bitwise branching of program transformation, the branching path can facilitate verification tools widen verification tasks over bitvector programs. Our experiment shows this exploitation of integer reasoning and interpolation enables competitive termination verification of bitvector programs and leads to the first effective technique for LTL verification of bitvector programs.


\end{abstract}
%

%
%

\section{Introduction}
\label{sec:motiv}

In many case, verification tools(e.g. $Ultimate$~\cite{heizmann_ultimate_nodate}) fail in termination and LTL verification tasks over bitvector programs, due to complex bitvector reasoning. In this work, we introduce two sets of program transforming rules namely bitwise branching, which transform a bitwise program to an over-approximated version by adding linear constraints, this enable verification tools verify bitvector programs in integer domain.

\added{For example,} while ranking functions \added{synthesis} is critical in proving termination and temporal properties of programs, it is challenging to \added{synthesize} ranking functions over the domain of bitvectors due to the complexity of bitvector logics ~\cite{kovasznai_complexity_2016}. $Ultimate$, a state-of-the-art verifier for LTL verification, over-approximates bitwise operators as uninterpreted functions, and returns $Unknown$ results for the following bitvector programs: 
\begin{center}
\begin{tabular}{l|l}
\hline
 {\bf (1) Termination} & {\bf (2) LTL} $\varphi = \GG(\FF(n < 0))$ \\
\hline
\hline
 \begin{lstlisting}[language=C,basicstyle=\tt\scriptsize]
 a = *; assume(a>0);
 while (x>0){
  a--;
  x = x & a;
}
\end{lstlisting} &
\begin{lstlisting}[language=C,basicstyle=\tt\scriptsize,escapechar=@]
while(1) {
  n = *; x = *; y = x-1;
  while (x>0 && n>0) {
    n++;
    y = x | n;
    x = x - y;
  }
  n = -1;
}
\end{lstlisting}
 \\
\hline
\end{tabular}
\end{center}
\noindent

In this work, we show that the key benefits of bitwise branching 
arise when concerned with termination and LTL.
Example {\bf (1)} involves a simple loop, in which $a$ is decremented, but the loop condition is on variable $x$, whose value is a bitvector expression over $a$.

Critical to verifying termination of this program are (1) proving the invariant($\mathcal{I}$) $ \code{x} > 0 \wedge \code{a} > 0$ 
within the body of the loop and (2) synthesizing a rank function. To prove the invariant, tools must show that it holds after a step of the loop's transition relation $T = x {>} 0 \wedge a' {=} a {-} 1 \wedge x' {=} x \code{\&} a'$, which requires reasoning about the bitwise-\code{\&} operation because if we simply treat the \code{\&} as an uninterpreted function, 
$\mathcal{I} \wedge T \wedge x' {>} 0 \centernot\implies \mathcal{I'}$. 
 
The bitwise branching strategy we introduce in \added{Sec.}~\ref{sec:bitwise} helps the verifier infer these invariants (and later synthesize rank functions) by transforming the bitvector assignment to $\code{x}$ into linear constraint $x \leq a$, but only under the condition that $x \geq 0$ and ${a \geq 0}$.

\begin{wrapfigure}[6]{r}[4pt]{5cm}
\vspace{-8mm}
\begin{lstlisting}[language=C,basicstyle=\tt\scriptsize,escapeinside={(*@}{@*)}]
  a = *; assume(a > 0);
  while (x > 0) {
    (*@\textcolor{blue}{ \{ x > 0 $\wedge$ a > 0 \} }@*)
    a--;
    (*@\colorbox{light-gray}{if (x >= 0 \&\& a >= 0)} @*)
    (*@\colorbox{light-gray}{then \{ x = *; assume(x <= a); \}}@*)
    (*@\colorbox{light-gray}{else \{}@*)x = x & a;(*@\colorbox{light-gray}{\}}@*)
  }
\end{lstlisting}
\end{wrapfigure}

In this case, bitwise branching translates the loop in Example {\bf (1)} as depicted in the gray box to the right. This transformation changes the transition relation of the loop body from $T$ (the original program) to $T'$:
\begin{gather*}
T' =  x {>} 0 \wedge a' {=} a {-} 1
          \wedge ((x {\geq} 0 \wedge a' {\geq} 0 \wedge x' {\leq} a')  \\
          \vee (\neg(x {\geq} 0 \wedge a' {\geq} 0) \wedge x' {=} x \code{\&} a'))
\end{gather*}
Importantly, when $\mathcal{I}$ holds, the else branch with the \code{\&} is infeasible, and thus we can treat the \code{\&} as an uninterpreted function and yet still prove that 
$\mathcal{I} \wedge T' \wedge x' {>} 0 \implies \mathcal{I'}$.
With the proof of $\mathcal{I}$ a tool can then move to the next step and synthesizing a ranking function $\mathcal{R}(x, a)$ that satisfies
$\mathcal{I} \wedge T' \implies \mathcal{R}(x, a) {\geq} 0 \wedge \mathcal{R}(x, a) {>} \mathcal{R}(x', a')$, namely,
$\mathcal{R}(x, a) = a$.

Bitwise branching also enables LTL verification of bitvector programs.
We also examine the bitwise  behavior of programs
such as Example {\bf (2)} above, with the LTL property $\GG(\FF(n < 0))$.
Our experiments show that with bitwise branching, our implementation
can prove LTL property of this program in 8.04s.

\section{Bitwise-branching}\label{sec:bitwise}

\newcommand\rr{\code{r}}
\newcommand\aaa{\code{a}}
\newcommand\bb{\code{b}}

We build our \emph{bitwise-branching} technique on the known strategy of transforming bitvector 
operations into integer approximations~\cite{MathSAT:Bitvector,CVC4:Int-blasting} but explore a new direction:
source-level transformations to introduce new conditional paths that approximate many (but not all) behaviors of a bitvector program. 
These new paths through the program have linear input conditions and linear output constraints and frequently 
cover all of the program's behavior (with respect to the goal property), but otherwise fall back on the original 
bitvector behavior when none of the input conditions hold.
We provide two sets of bitwise-branching rules:
\newcommand\ruleLabel[1]{}
\begin{figure}[!t]\centering
\subfloat[Rewriting rules for arithmetic expressions.]{\centering
  \label{fig:rewriting_rules}
  \begin{adjustbox}{max width=0.5\textwidth}
    \footnotesize
\begin{tabular}{rllll}

  \rewrite{e_1=0}{e_1 \texttt{\&} e_2}{0} & \ruleLabel{R-And-0}\\

  \rewrite{(e_1=0 \vee e_1=1) \wedge e_2=1}{e_1 \texttt{\&} e_2}{e_1} & \ruleLabel{R-And-1}\\

  \rewrite{(e_1=0 \vee e_1=1) \wedge (e_2=0 \vee e_2=1)}{e_1 \texttt{\&} e_2}{e_1 \texttt{\&\&} e_2} & \ruleLabel{R-And-LOG}\\
 
  \rewrite{e_1 \geq 0 \wedge e_2 = 1}{e_1 \texttt{\&} e_2}{e_1 \texttt{\%} 2} & \ruleLabel{R-And-LBS}\\
  
  \rewrite{e_2=0}{e_1 \texttt{|} e_2}{e_1} & \ruleLabel{R-Or-0}\\
  
  \rewrite{(e_1=0 \vee e_1=1) \wedge e_2=1}{e_1 \texttt{|} e_2}{1} & \ruleLabel{R-Or-1} \\

  \rewrite{e_2=0}{e_1 \texttt{\^{}} e_2}{e_1} & \ruleLabel{R-Xor-0}\\
 
  \rewrite{e_1=e_2=0 \vee e_1=e_2=1}{e_1 \texttt{\^{}} e_2}{0} & \ruleLabel{R-Xor-Eq}\\
  
  \rewrite{(e_1=1 \wedge e_2=0) \vee (e_1=0 \wedge e_2=1)}{e_1 \texttt{\^{}} e_2}{1} & \ruleLabel{R-Xor-Neq}\\

  \rewrite{e_1 \geq 0 \wedge e_2 = \texttt{CHAR\_BIT * sizeof}(e_1) - 1}{e_1 \texttt{>>} e_2}{0} & \ruleLabel{R-RightShift-Pos}\\
  
  \rewrite{e_1 < 0 \wedge e_2 = \texttt{CHAR\_BIT * sizeof}(e_1) - 1}{e_1 \texttt{>>} e_2}{-1} & \ruleLabel{R-RightShift-Neg}

\end{tabular}

   \end{adjustbox}
}
\vspace{5pt}
  \hrule
\vspace{5pt}
\subfloat[Weakening rules for relational expressions and assignments.\\$\texttt{op}_{le} \in \{\texttt{<,<=,==,:=}\}$, $\texttt{op}_{ge} \in \{\texttt{>,>=,==,:=}\}$, and $\texttt{op}_{eq} \in \{\texttt{==,:=}\}$
]{
    \label{fig:weakening_rules}
    \begin{adjustbox}{max width=0.5\textwidth}
      \footnotesize
\begin{tabular}{rllll}
  \weaken{e_1 \geq 0 \wedge e_2 \geq 0}{r ~\texttt{op}_{le}~ e_1 \texttt{\&} e_2}{r \texttt{<=} e_1 ~\texttt{\&\&}~ r \texttt{<=} e_2} & \ruleLabel{W-And-Pos}\\

  \weaken{e_1 < 0 \wedge e_2 < 0}{r ~\texttt{op}_{le}~ e_1 \texttt{\&} e_2}{r \texttt{<=} e_1 ~\texttt{\&\&}~ r \texttt{<=} e_2 ~\texttt{\&\&}~ r \texttt{<} 0} & \ruleLabel{W-And-Neg}\\  
  
  \weaken{e_1 \geq 0 \wedge e_2 < 0}{r ~\texttt{op}_{eq}~ e_1 \texttt{\&} e_2}{0 \texttt{<=} r ~\texttt{\&\&}~ r \texttt{<=} e_1} & \ruleLabel{W-And-Mix}\\ 
  
  \weaken{(e_1=0 \vee e_1=1) \wedge (e_2=0 \vee e_2=1)}{(e_1 \texttt{|} e_2) \texttt{==} 0}{e_1 \texttt{==} 0 \texttt{ \&\& } e_2 \texttt{==} 0} & \ruleLabel{R-Or-LOG}\\
  
  \weaken{e_1 \geq 0 \wedge \texttt{is\_const}(e_2)}{r ~\texttt{op}_{ge}~ e_1 \texttt{|} e_2}{r \texttt{>=} e_2} & \ruleLabel{W-Or-Const}\\ 
  
  \weaken{e_1 \geq 0 \wedge e_2 \geq 0}{r ~\texttt{op}_{ge}~ e_1 \texttt{|} e_2}{r \texttt{>=} e_1 \texttt{ \&\& } r \texttt{>=} e_2} & \ruleLabel{W-Or-Pos}\\
  
  \weaken{e_1 < 0 \wedge e_2 < 0}{r ~\texttt{op}_{eq}~ e_1 \texttt{|} e_2}{r \texttt{>=} e_1 \texttt{ \&\& } r \texttt{>=} e_2 \texttt{ \&\& } r \texttt{<} 0} & \ruleLabel{W-Or-Neg}\\  
  
  \weaken{e_1 \geq 0 \wedge e_2 < 0}{r ~\texttt{op}_{eq}~ e_1 \texttt{|} e_2}{e_2 \texttt{<=} r \texttt{ \&\& } r \texttt{<} 0} & \ruleLabel{W-Or-Mix}\\ 
  
  \weaken{e_1 \geq 0 \wedge e_2 \geq 0}{r ~\texttt{op}_{ge}~ e_1 \texttt{\^{}} e_2}{r \texttt{>=} 0} & \ruleLabel{W-XOr-Pos}\\
  
  \weaken{e_1 < 0 \wedge e_2 < 0}{r ~\texttt{op}_{ge}~ e_1 \texttt{\^{}} e_2}{r \texttt{>=} 0} & \ruleLabel{W-XOr-Neg}\\  
  
  \weaken{e_1 \geq 0 \wedge e_2 < 0}{r ~\texttt{op}_{le}~ e_1 \texttt{\^{}} e_2}{r \texttt{<} 0} & \ruleLabel{W-XOr-Mix}\\
  
  \weaken{e_1 \geq 0}{r ~\texttt{op}_{le}~ {\sim}e_1}{r \texttt{<} 0} & \ruleLabel{W-Cpl-Pos}\\
  
  \weaken{e_1 < 0}{r ~\texttt{op}_{ge}~ {\sim}e_1}{r \texttt{>=} 0} & \ruleLabel{W-Cpl-Neg}\\
  
\end{tabular}
\end{adjustbox}
}
\caption{Rewriting rules. Commutative closures omitted for brevity.}
\vspace{-6mm}
\end{figure}

{\bf 1. Rewriting rules} of the form $\inlrewrite{\mathcal{C}}{e_{bv}}{e_{int}}$ in Fig. \ref{fig:rewriting_rules}. These rules are applied to bitwise arithmetic expressions $e_{bv}$ and specify a condition $\mathcal{C}$ for which one can use integer approximate behavior $e_{int}$ of $e_{bv}$.
 
In other words, rewriting rule $\inlrewrite{\mathcal{C}}{e_{bv}}{e_{int}}$ can be applied only when $\mathcal{C}$ holds and a bitwise arithmetic expression $e$ in the program structurally matches its $e_{bv}$ with a substitution $\delta$. Then, $e$ will be transformed into a conditional approximation: $\mathcal{C}\delta ~?~ e_{int}\delta : e_{\added{bv}}$.
Note that, although modulo-2 is computationally more expensive, it is often more amenable to integer reasoning strategies.
  For conciseness, we omitted variants that arise from commutative re-ordering of the rules (in both Figs.~\ref{fig:rewriting_rules} and~\ref{fig:weakening_rules}). 


{\bf 2. Weakening rules} of the form $\inlweaken{\mathcal{C}}{s_{bv}}{s_{int}}$ are in Fig. \ref{fig:weakening_rules}. These rules are applied to relational condition expressions (\eg~from assumptions) and assignment statements $s_{bv}$, specifying an integer condition $\mathcal{C}$ and over-approximation transition constraint $s_{int}$.
  

When the rule is applied to a statement (as opposed to a conditional),  replacement $s_{int}$ can be implemented as $assume(s_{int})$.
 When a weakening rule $\inlweaken{\mathcal{C}}{s_{bv}}{s_{int}}$ is applied to an assignment $s$ with substitution $\delta$, the transformed statement is $\texttt{if } \mathcal{C}\delta \texttt{ assume(} s_{int}\delta \texttt{) else } s_{\added{bv}}$.
In addition, when $s_{bv}$ of a weakening rule can be matched to the condition $\texttt{c}$ in an
$\texttt{assume(c)}$ of the original program via a substitution $\delta$, then the $\texttt{assume(c)}$ statement is transformed to
$\texttt{if } \mathcal{C}\delta \texttt{ then assume(} s_{int}\delta \texttt{) else assume(c)}$.

The rules in Fig.~\ref{fig:rewriting_rules} and Fig.~\ref{fig:weakening_rules} were developed empirically, from the 
reachability/termination/LTL benchmarks
, especially, based on patterns found in decompiled binaries(enable us to verify properties over binaries). We then generalized these rules to expand coverage, proofs for each rule were done with Z3.

\section{Experiments \& Conclusion}
\label{sec:eval}

\medskip\noindent

We implemented bitwise branching via a translation algorithm, in a fork of $Ultimate$, which is the state-of-the-art LTL prover and the only mature LTL verifier that supports bitvector programs, we denote our branch $UltimateBwB$. 
To our knowledge, there are no available bitwise benchmarks with LTL properties so we create new benchmarks for this purpose:
(i) New hand-crafted benchmarks called $LTLBit$ of 42 C programs with LTL properties, in which bitwise operations are heavily used in assignments, loop conditions, and branching conditions. There are 22 programs in which the provided LTL properties are satisfied (\OK) and 20 programs in which the LTL properties are violated
(\NOK).
(ii) Benchmarks adapted from the $BitHacks$ programs ~\cite{bithacks}, consisting of 26 programs with LTL properties (18 satisfied and 8 violated).


\begin{table} 
  \caption{LTL Verification over Bitvector programs resutls}
  \label{ltlEval}
\renewcommand{\header}[1]{\parbox{5em}{\centering #1}}
\newcolumntype{Y}{>{\centering\arraybackslash}X}
\begin{tabularx}{0.5\textwidth}{l *{4}{Y}}
      \toprule
            & \multicolumn{2}{c}{(i)$BitHacks$} & \multicolumn{2}{c}{\header{(ii)$LTLBit$}}                                               \\
      \cmidrule(r){2-3}\cmidrule{4-5}
           & $Ultimate$  & $w.BwB$ & $Ultimate$ & $w.BwB$ \\
      \cmidrule(r){2-2}\cmidrule(r){3-3}\cmidrule(r){4-4}\cmidrule{5-5}
      \OK   & 3                            & 10                                & -                   & 21                \\
      \NOK  & -                            & 7                                 & -                   & 20                \\
      \UNK(Unknown)  & 21                  & 5                                 & 42                  & -                 \\
      \TOUT(TimeOut) & 1                   & 1                                 & -                   & 1                 \\
      \OOM(Out of Memory)  & 1             & 3                                 & -                   & -                 \\
      \bottomrule
\end{tabularx}

\vspace{-6mm}
\end{table}
Table~\ref{ltlEval} summarizes the result of applying $Ultimate$ and $UltimateBwB$ on these two bitvector
benchmarks.
$UltimateBwB$ outperforms $Ultimate$ thanks to our newly added abstraction of bitwise operations. 
$UltimateBwB$ can successfully verify 41 of 42 programs in $LTLBit$ and 18 of 26 BitHacks programs while $Ultimate$ can only handle a few of them.
Note that we have more out-of-memory results in BitHacks Benchmarks, perhaps due to memory consumption reasoning about the introduced paths.

In conclusion, bitwise branching appears to be the first effective technique for verifying LTL properties of bitvector programs.

\bibliographystyle{IEEEtran}
\bibliography{biblio}

\end{document}